\documentclass[12pt]{iopart}

\usepackage{graphicx, url}
\usepackage{amssymb}
\begin{document}

\title{Triangular clustering in document networks}

\author{Xue-Qi Cheng$^{1}$\footnote{Author to whom any correspondence should be addressed.}, Fu-Xin Ren$^{1}$,
Shi Zhou$^{2}$ and Mao-Bin Hu$^{3}$}

\address{$^{1}$ Institute of Computing Technology, Chinese Academy of Sciences, Beijing, 100190,
P.R.~China}
\address{$^{2}$ Department of Computer Science, University College London, Malet Place, London, WC1E 6BT, United Kingdom}
\address{$^{3}$ School of Engineering Science, University of Science and
Technology of China, Hefei, 230026, P.R.~China} \eads{\mailto{cxq@ict.ac.cn},
\mailto{renfuxin@software.ict.ac.cn}, \mailto{s.zhou@adastral.ucl.ac.uk},
\mailto{humaobin@ustc.edu.cn}}

\date{\today}

\begin{abstract}
Document networks are characteristic in that a document node, e.g.~a
webpage or an article, carries meaningful content. Properties of
document networks are not only affected by topological connectivity
between nodes, but also strongly influenced by the semantic relation
between content of the nodes. We observe that document networks have
a  large number of triangles and a high value of clustering
coefficient. And there is a strong correlation between the
probability of formation of a triangle and the content similarity
among the three nodes involved. We propose the degree-similarity
product (DSP) model which well reproduces these properties. The
model achieves this by using a preferential attachment mechanism
which favours the linkage between nodes that are both popular and
similar. This work is a step forward towards a better understanding
of the structure and evolution of document networks.

\end{abstract}


\maketitle

\section{Introduction}

In recent years studying the structure, function and evolution of
complex networks in society and nature has become a major research
focus~\cite{Watts1998,Barab'asi1999,Albert2002,Newman2003}. Examples
of complex networks include the Internet, the World Wide Web, the
international aviation network, social collaborations between a
group of people,  protein interactions in a cell, to name just a
few.  These networks exhibit a number of interesting properties,
such as  short average distance between a pair of nodes in
comparison with large network size~\cite{Watts1998}, the clustering
structure where one's friends are friends of each other, and the
power law distribution of the number of connections a node
has~\cite{Barab'asi1999}.

This paper concerns one particular type of complex networks, the
document networks, such as the Web and the citation networks.
Document networks are characteristic in that a document node, e.g.~a
webpage or an article, carries text or multimedia content.
Properties of document networks are not only affected by topological
connectivity between nodes, but also strongly influenced by semantic
relation between the content of nodes. Research on document networks
is relevant to a number of issues, such as the Web navigation and
information retrieval~\cite{Kleinberg2001,Menczer2002,Fenner2006}.

Menczer~\cite{Menczer2004} reported that the  probability of linkage
between two documents increases with the similarity between their
content. Based on this observation, he proposed the
degree-similarity mixture (DSM) model, which successfully reproduces
two important properties of document networks: the power-law
connectivity distribution and the increasing linkage probability as
a function of content similarity. The DSM model remains one of the
most advanced models for document networks.
%

Recently we reported that document networks exhibit a number of
triangular clustering properties, for example they have huge numbers
of triangles  and high clustering coefficients, and  there is a
positive relation between the probability of formation of a triangle
and the content similarity among the three documents
involved~\cite{Cheng2007}. Menczer's DSM model focuses on the
connectivity and content properties between two nodes, and it
produces only around 5\% of triangles in real document networks.
There are a number of topology models which can produce networks
with a power-law distribution of connectivity  with high clustering
coefficient, such as a network  model
in~\cite{Toivonen2006,Kumpula2007} which is based on the balance
between different types of attachment mechanisms, i.e.~cyclic
closure and focal closure. This model, however, do not has the
ingredient of document content in its generative mechanisms and can
not reproduce content-related properties of document networks.

In this paper, we examine and model the triangular clustering
properties of document networks. In Section~\ref{sec:2}, we firstly
introduce two datasets of real document networks, we then define a
number of metrics to quantify connectivity and content properties,
and finally we review Menczer's DSM model. In Section~\ref{sec:3} we
propose our degree-similarity product (DSP) model, where a node's
ability of acquiring a new link is given as a \emph{product}
function of node connectivity and content similarity between nodes.
In Section~\ref{sec:4} we evaluate our DSP model against the real
data and show that the model reproduces not only the connectivity
and content properties between two nodes, but also the triangular
clustering properties involving three nodes. In Section~\ref{sec:5}
we conclude the paper.

\section{Triangular clustering in document networks}
\label{sec:2}
\begin{table}\caption{\label{tab:modelEvaluation}
Evaluation of the degree-similarity mixture (DSM) model and the
degree-similarity product (DSP) model against the {WT10g} data and
the {PNAS} data, respectively. Topological properties shown are the
number of nodes $N$, the number of links $L$, the total number of
weak triangles $\triangle$ and the average clustering
coefficient~$\langle C\rangle$. For each model, ten networks are
generated for the {WT10g} data and the {PNAS} data respectively, and
results are averaged.} \footnotesize\rm
\renewcommand{\arraystretch}{1.3}
\begin{tabular*}{\textwidth}{@{\extracolsep{\fill}}c|ccc}
\br
~~~~Properties~~  &WT10g   & DSM Model &  DSP Model \\
\hline
$N$& 50~000   & 50~000   &  50~000\\
$L$ & 233~692  & 233~692  &  234~020$_{\pm1228}$\\
$\triangle$  & 1266~730  & 62~503$_{\pm187}$ &  1233~308$_{\pm18~467}$\\
$\langle C\rangle$ & 0.153  & 0.062$_{\pm0.001}$ & 0.121$_{\pm0.001}$\\
\br
~~~~Properties~~  &PNAS   & DSM Model &  DSP Model \\
\hline
$N$&  28~828  & 28~828 &28~828\\
$L$ &  40~610 & 40~610 & 40 580$_{\pm215}$\\
$\triangle$  &  13~544  & 868$_{\pm24}$ & 13~583$_{\pm329}$\\
$\langle C\rangle$ &  0.214 & 0.021$_{\pm0.0002}$ & 0.139$_{\pm0.001}$\\
\br
\end{tabular*}
\end{table}

\subsection{Two Datasets}
In this study we examine the following two datasets of real document networks.

\begin{itemize}

\item \texttt{WT10g} data, which is a webpage network where a webpage is a node and two webpages
are connected if there is a hyperlink between them. The WT10g data
are proposed by the annual international Text REtrieval Conference
(\url{http://trec.nist.gov}) and distributed by CSIRO
(\url{http://es.csiro.au/TRECWeb}). The data preserve properties of
the Web and have been widely used in research on information
modelling and retrieval~\cite{Soboroff2002,Chiang2005}. The data
contain $1.7$ million webpages, hyperlinks among them and the text
content on each webpage. We study ten randomly sampled subsets of
the WT10g data. Each subset contains $50,000$ webpages with the URL
domain name of \emph{.com}. (A recent study has shown that subsets
sampled from different or mixed domains exhibit similar
properties~\cite{Cheng2007}.) Observations in this paper are
averaged over the ten subsets.

\item \texttt{PNAS} data, which is a citation network where an article is a node and two article
are linked if they have a citation relation. It contains $28,828$
articles published by the Proceedings of the National Academy of
Sciences (PNAS) of the United States of America from $1998$ to
$2007$. We crawled the data at the journal's website
(\url{http://www.pnas.org}) in May 2008 and used each article's
title and abstract as its content.

\end{itemize}

\subsection{Triangle and Clustering Coefficient}

Triangle is the basic unit for clustering structure and  network
redundancy~\cite{Cheng2007,Schank2005, Serrano2006, Fagiolo2007,
Arenas2008, Zhou2007}. Triangle-related properties have been used to
quantify network transitivity~\cite{Schank2005} and characterise the
structural invariance across web sites~\cite{Zhou2007}.

The most widely studied triangle-related property is the clustering
coefficient, $C$, which measures how tightly a node's neighbours are
interconnected with each other~\cite{Watts1998,Newman2003}.
Clustering coefficient is calculated as the ratio of the number of
triangles formed by a node and its neighbours to the maximal number
of triangles they can have. When $C=1$ a node and its neighbours are
fully interconnected and form a clique; and when $C=0$ the
neighbours do not know each other at all. The average clustering
coefficient over all nodes measures the level of clustering
behaviour in a network.

Note that triangle and clustering coefficient are not trivially
related. As shown in Table~\ref{tab:modelEvaluation}, the total
number of triangles, $\triangle$, in the WT10g data is almost 100
times of that in the PNAS data. The density of triangles in the
WT10g data, measured by $\triangle/N$ or $\triangle/L$, is also many
times larger. However the average clustering coefficient, $\langle
C\rangle$, of the WT10g data is smaller than that of the PNAS data.

\subsection{Content Similarity and Linkage Probability}

\begin{figure}
\includegraphics[width=8cm]{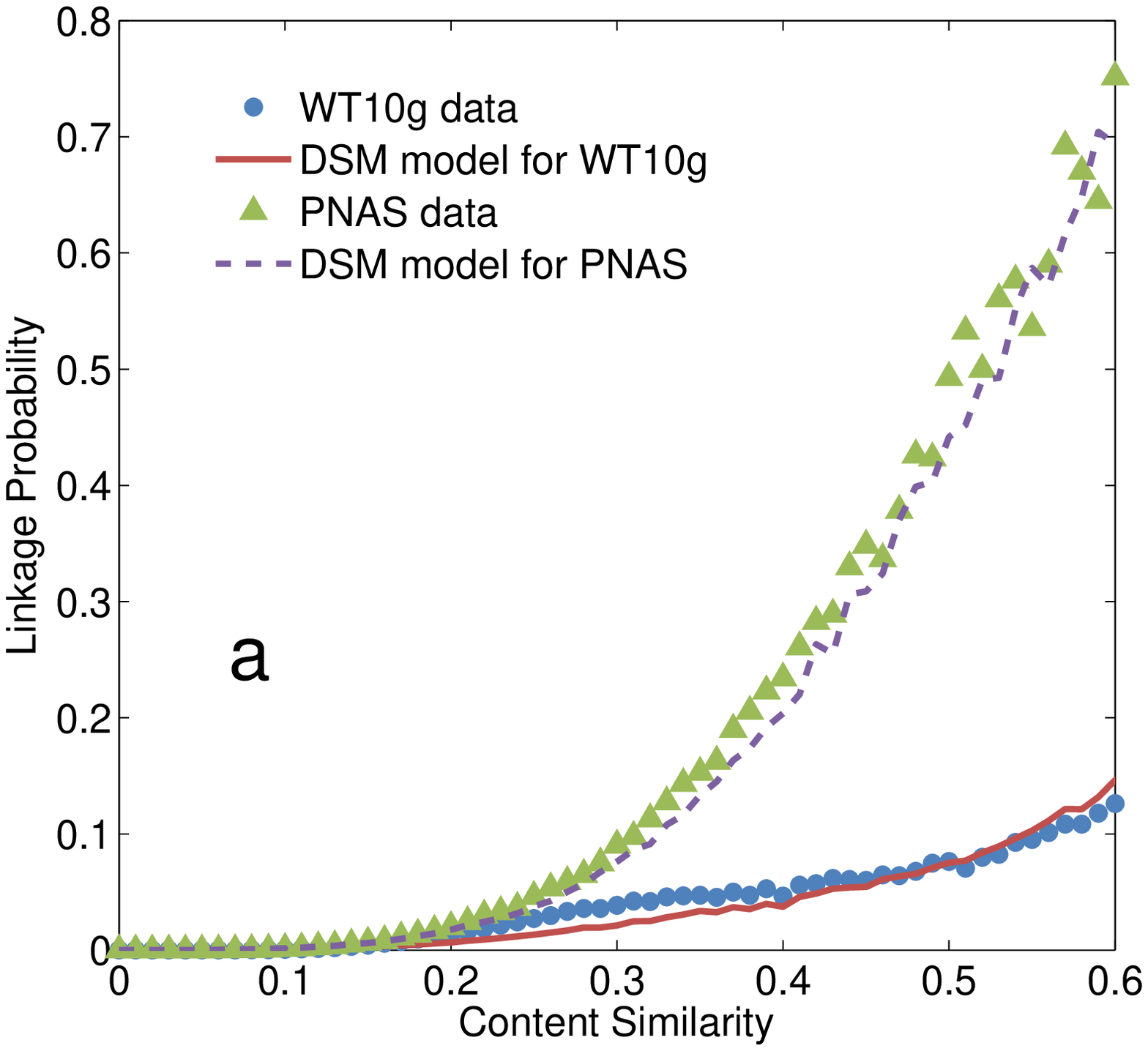}
\includegraphics[width=8cm]{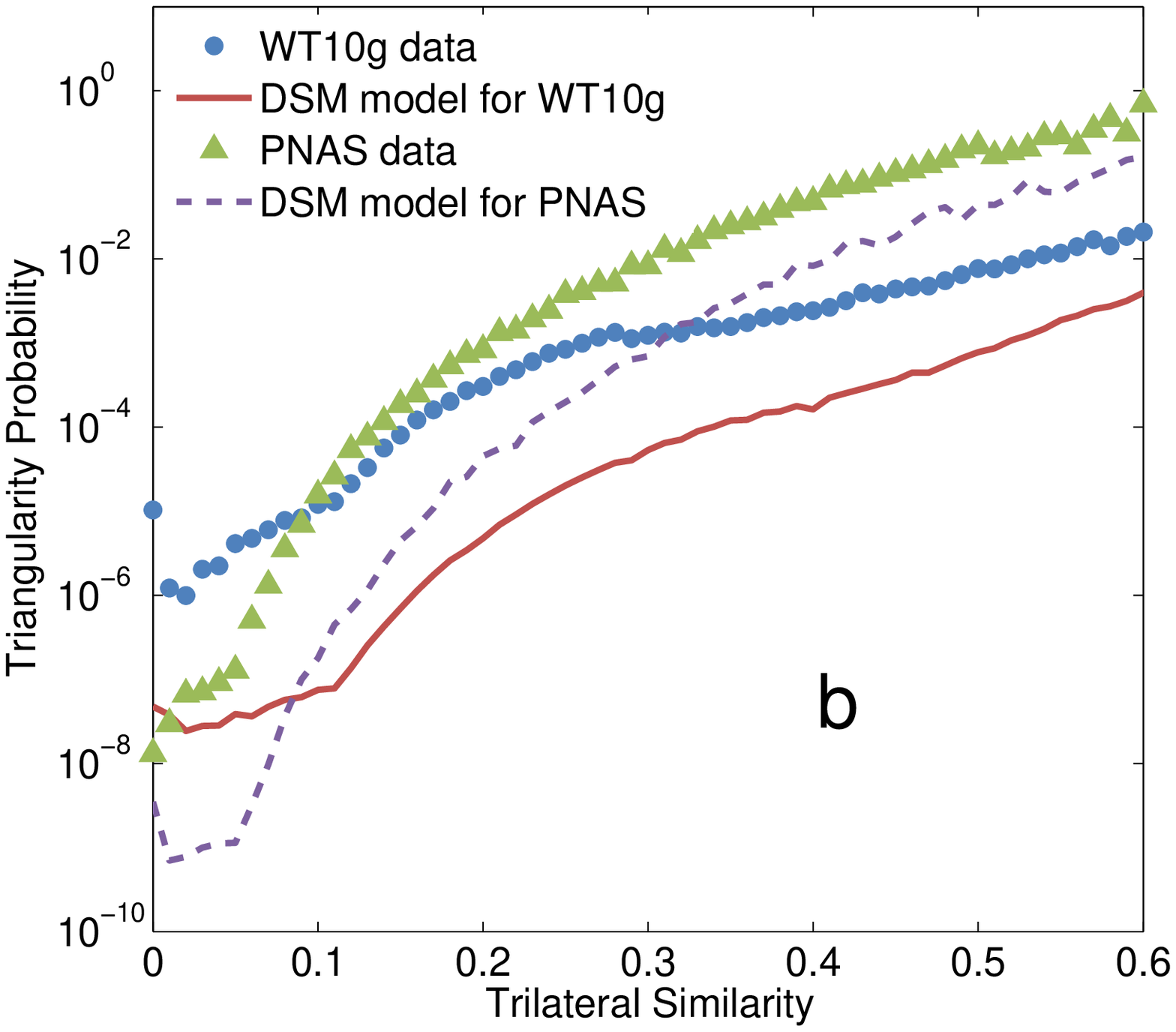}
\caption{\label{fig:DSM} Linkage probability and triangularity
probability for the WT10g webpage network and the PNAS citation
network. The results are compared with Menczer's DSM model.
(a)~Linkage probability $P(R)$  as a function of content similarity
$R$. (b)~Triangularity probability $P(R^{\triangle})$, in
logarithmic scale, as a function of trilateral similarity
$R^{\triangle}$.}
\end{figure}

For a given document network, we collect keywords present in all
documents in the network and construct a keyword vector
space~\cite{Salton1989, Ricardo1999}. The content of a document is
then represented as a keyword vector, $\overrightarrow{X}$, which
gives the frequency of each keyword's appearance in the document.
The content similarity, or relevance, $R$, between two documents,
$i$ and $j$, is quantified by the cosine of their vectors:
\begin{equation}
R_{ij}={R_{ji}}=\frac{\|\overrightarrow{X_i} \bullet
\overrightarrow{X_j}\|}{\|\overrightarrow{X_i}\| \cdot \|\overrightarrow{X_j}\|}.
\label{eq:cos}
\end{equation}
When $R_{ij}=1$ the content of the two documents are highly related
or similar; when $R_{ij}=0$ the two documents have very little in
common.  The linkage probability, $P(R)$, is the probability that
two nodes with content similarity $R$ are connected in the network.
It is calculated as $P(R)=M^*(R)/M(R)$, where $M(R)$ is the total
number of node pairs (connected or not) whose content similarity is
$R$, and $M^*(R)$ is the number of such node pairs which are
actually connected in the network.

Figure~\ref{fig:DSM}(a) shows that in document networks the linkage
probability increases with the content similarity, i.e.~the more
similar the more likely two documents are connected. For example in
the PNAS citation network, if two articles have $R=0.5$ there is a
50\% chance that they have a citation relation, by comparison the
chance is very low when $R<0.2$.

\subsection{Trilateral Similarity and Triangularity Probability}

In document networks, if a node is similar to a second node and this
second node is similar to a third node, then the first and third
nodes are also similar. Here we define a new metric called the
\emph{trilateral similarity}, $R^{\triangle}$, which measures the
minimum content similarly among three nodes. For three document
nodes $i$, $j$ and $k$, the trilateral similarity is the smallest
(bilateral) content similarity between each pair of the three nodes,
i.e.
\begin{equation}
R^{\triangle}_{ijk}=\min\{R_{ij},R_{ik},R_{jk}\}.
\end{equation}
Similarly we define the triangularity probability,
$P(R^{\triangle})$, as the probability that three nodes with the
trilateral similarity $R^{\triangle}$ form a triangle. In this study
we consider weak triangles, each of which is a circle of three nodes
with at least one link (at any direction) between each pair  of the
three nodes.

Figure~\ref{fig:DSM}(b) shows that the triangularity probability is
sensitive to the trilateral similarity. When the trilateral
similarity $R^{\triangle}$ increases from $0.1$ to $0.5$, the
triangularity probability increases two orders of magnitude for the
WT10g data and four orders of magnitude for the PNAS data,
respectively.

We note that for a given value of content similarity or trilateral
similarity, the cube of the (bilateral) linkage probability provides
the lower bound of the triangularity probability. But these two
quantities are not trivially related because the later is strongly
determined by a network's triangular clustering structure.

\subsection{Degree-Similarity Mixture (DSM) Model}
The degree-similarity mixture (DSM) model was introduced by Menczer
in 2004~\cite{Menczer2004}. The model's generative mechanism
incorporates content similarity in the formation of document links.
At each step, one new document is added and attached by $m=L/N$ new
links to existing documents. At time step $t$, the probability that
the new document $t$ is attached to the existing document $i$ is

\begin{equation}
Pr(i)=\alpha {k_i \over mt}+(1-\alpha)\overline{Pr}(i),
~~\overline{Pr}(i) \propto ({1 \over R_{it}}-1)^{-\gamma},
\label{eq:DSM}
\end{equation}
where $i < t$; $k_i$ is the number of connections, or degree, of
node $i$; $R$ is calculated from document content of the given
network; $\gamma$ is a constant which is calculated based on real
data; and $\alpha$ is a preferential attachment parameter. The first
term of Equation~(\ref{eq:DSM}) favours an old node which is already
well connected and the second term favours one whose content is
similar to the new node. The tunable parameter
$0\leqslant\alpha\leqslant 1$ models the balance between choosing a
popular node with large degree or choosing a similar node with high
content similarity.

\begin{table}\caption{\label{tab:parameters}
Parameters used by the two models for the datasets. } \footnotesize\rm
\renewcommand{\arraystretch}{1.3}
\begin{tabular*}{\textwidth}{@{\extracolsep{\fill}}c|cc}
\br
DSM Model parameters   & WT10g   & PNAS \\
\hline
$\alpha$ &0.1  &0.01  \\
$\gamma$ &3.5  & 3.5 \\
 \br
DSP Model parameters   & WT10g   & PNAS \\
\hline
 $\beta_1$ & 5 & 7\\
 $\beta_2$ & 1 & 4 \\
 $\alpha$  & $10^{-12}$ & $10^{-12}$  \\
 $\lambda$ & 6 & 8  \\
\br
\end{tabular*}
\end{table}

For each of the two document networks under study, we use the DSM
model to grow ten networks to the same size of the real network  and
results are averaged over the ten networks (see
Table~\ref{tab:modelEvaluation}). Table~\ref{tab:parameters} gives
the model parameters which are obtained, as
Menczer~\cite{Menczer2004} did, by best fitting. Menczer has shown
that the DSM model is able to reproduce the degree distribution of
document networks. Figure~\ref{fig:DSM}(a) shows the DSM model also
produces a sound prediction on the relation between linkage
probability and content similarity.

In terms of triangular clustering properties,
Table~\ref{tab:modelEvaluation} shows that the model, however,
produces only around $5\%$ of the total number of triangles
contained in the real networks and underestimates the average
clustering coefficient of the networks. Figure~\ref{fig:DSM}(b)
shows the model also significantly underestimates the correlation
between triangularity probability and trilateral similarity.

\section{Degree-Similarity Product (DSP) model}
\label{sec:3}

In this paper we introduce a new generative model for document networks, we call it the
degree-similarity product (DSP) model. Our model is partially inspired by the multi-component graph
growing models of~\cite{Callaway2001, Krapivsky2002}. The model starts from an initial seed of a
pair of linked nodes. At each time step, one of the following two actions is taken:
\begin{itemize}
\item Growth: with probability $p$, a new isolated node is introduced to the network. Parameter $p$
is a constant, which is given by the numbers of nodes and links of the generated network,
$p=N/(N+L)$, and determines the average node degree of the generated network,
i.e.\,$<k>=2L/N=2(1-p)/p$.

\item DSP preferential attachment: with probability $(1-p)$, a new link is attached between two
nodes. The link starts from node $i$ and ends at node $j$. The two nodes are chosen by the
following preferential probabilities:
\begin{equation}
\Pi(i)={{k_{i}^{out} + \beta_{1}}\over{\sum_{m}}(k_{m}^{out} + \beta_{1})},
\label{eq:orignal node}
\end{equation}
\begin{equation}
\Pi_i(j)=\frac{(k_{j}^{in} + \beta_{2})(R_{ij}^{\lambda}+\alpha)}{\sum_{l}\left[(k_{l}^{in} +
\beta_{2})(R_{il}^{\lambda}+\alpha)\right]}, \label{eq:target node}
\end{equation}
where $k_{i}^{out}$ is the out-degree of node $i$, $k_{j}^{in}$ is
the in-degree of node $j$, $m$ and $l$ run over all existing nodes,
$l\neq i$. The content similarity $R_{ij}$ is calculated from
document content of the given network. Parameters $\beta_1$,
$\beta_2$, $\alpha$ and $\lambda$ all take positive values.
$\beta_1$ and $\beta_2$ give nodes with $k^{out}=0$ or $k^{in}=0$,
respectively, an initial ability of acquiring links. $\alpha$ allows
that even very different documents (with $R\simeq 0$) still have a
chance to link with each other. $\lambda$ tunes the weight of the
content similarity in choosing a link's ending node.
\end{itemize}

It is notable that Equation~\ref{eq:target node} is a product
function of degree and content similarity. This ensures that links
are preferentially attached between nodes which are \emph{both}
popular \emph{and} similar. As shown in the following section, this
mechanism effectively increases the chance of forming triangles
among similar nodes.

\section{Evaluation of DSP Model}
\label{sec:4}

For each of the two document networks, we generate ten networks using the DSP model with different
random seeds. We avoid creating self-loops and duplicate links. The ten networks are grown to the
same size as the target network. Results are then averaged over the ten networks.

\begin{figure*}
\includegraphics[width=8cm]{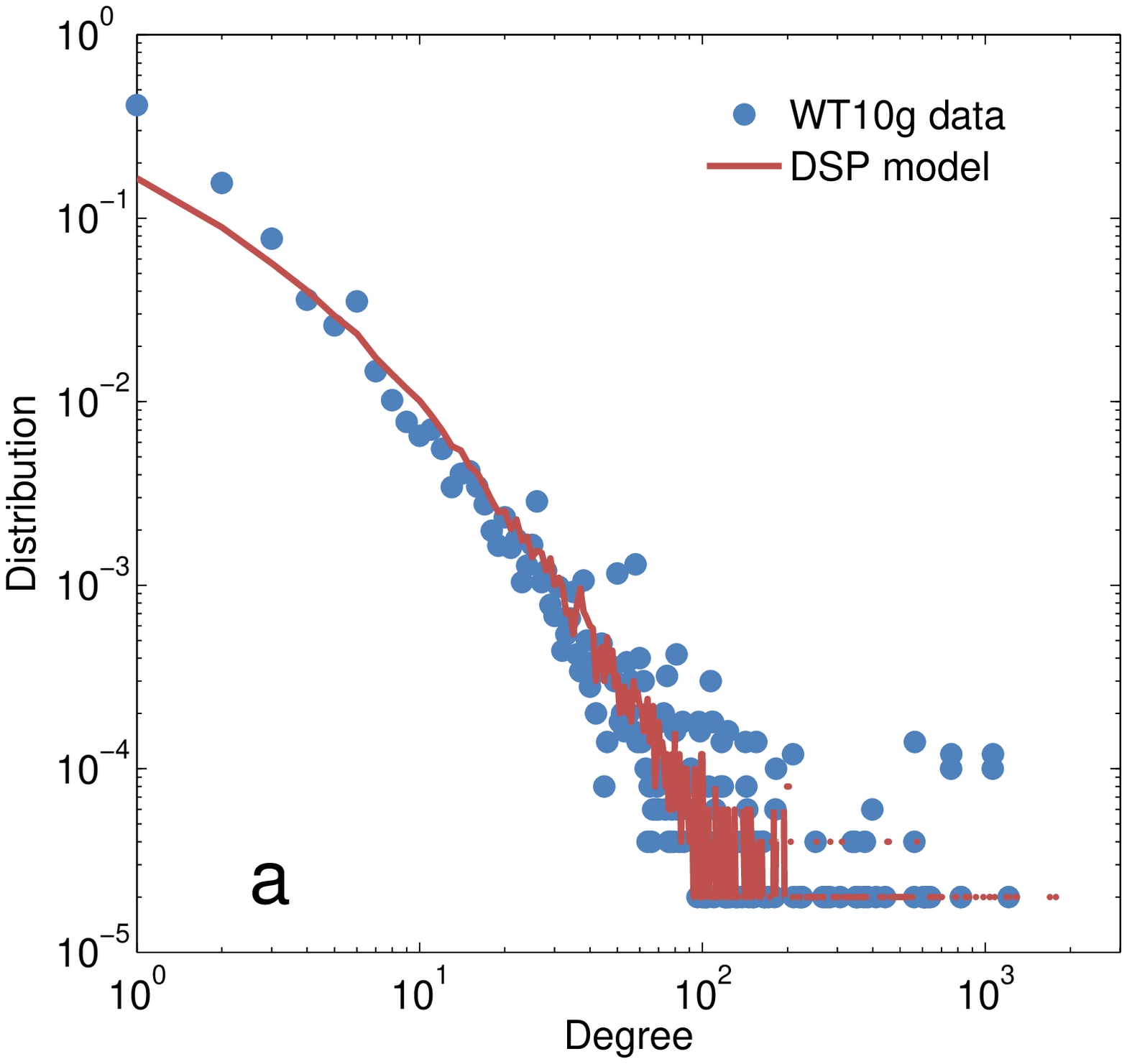}
\includegraphics[width=8cm]{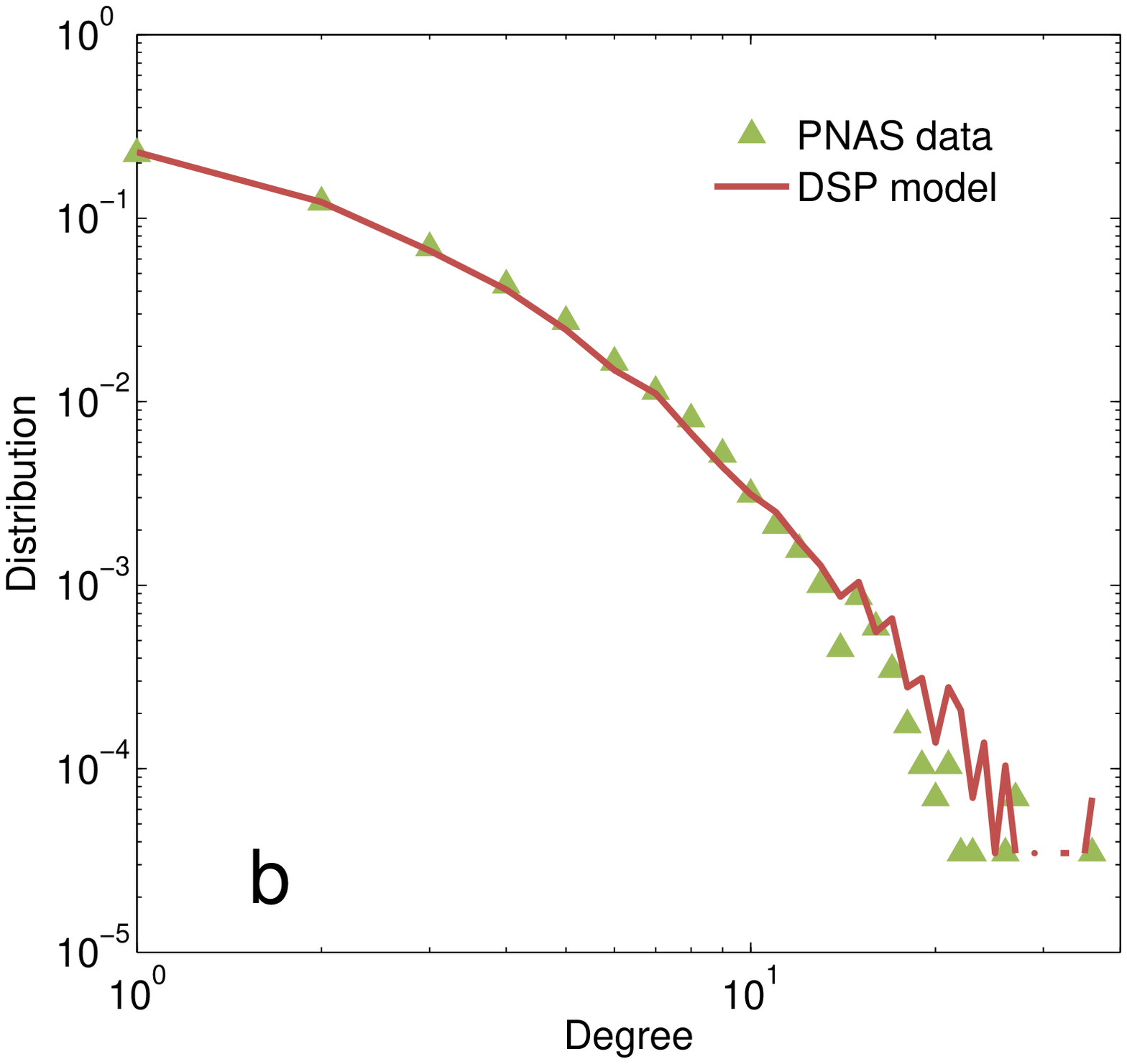}
\includegraphics[width=8cm]{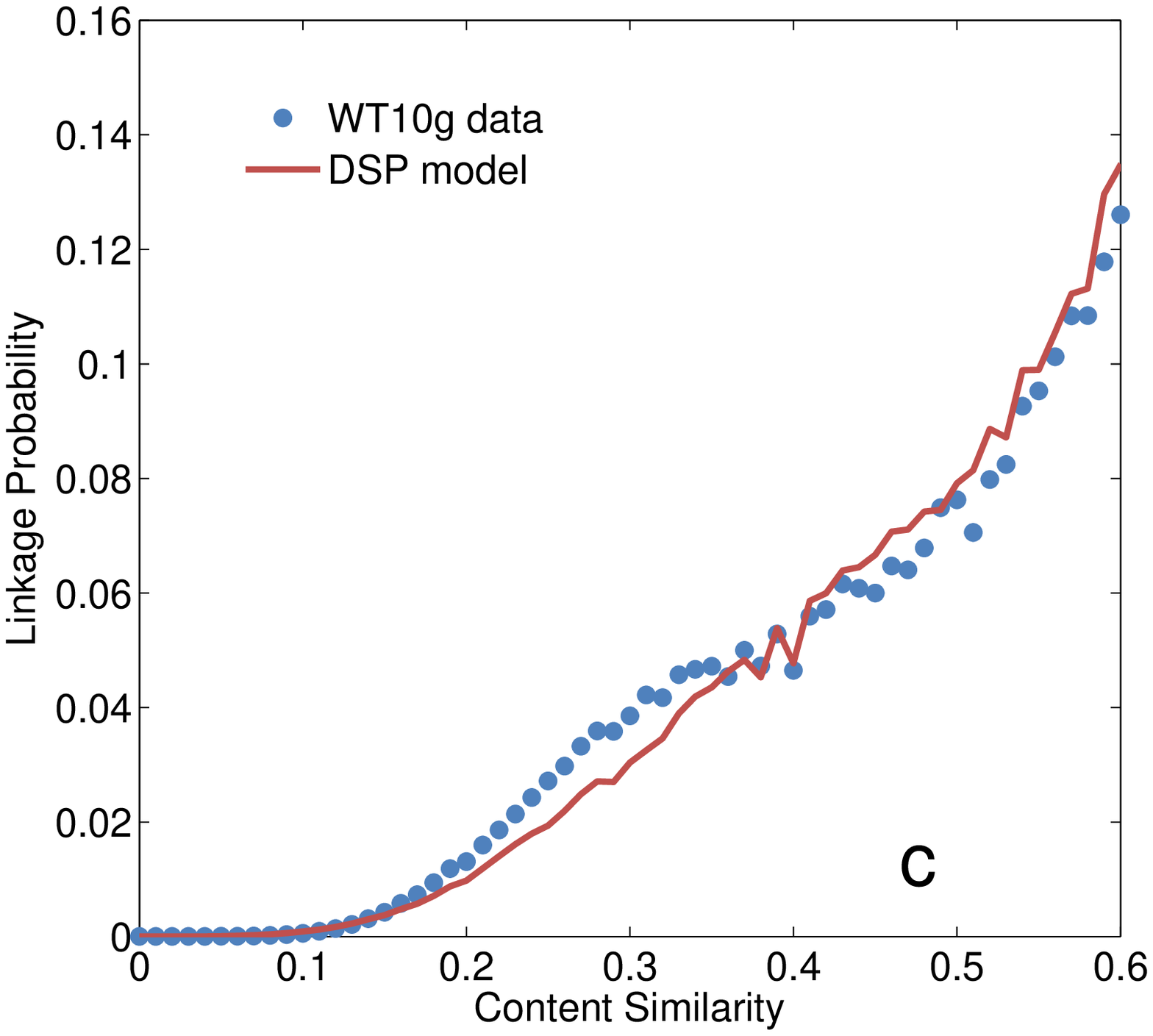}
\includegraphics[width=8cm]{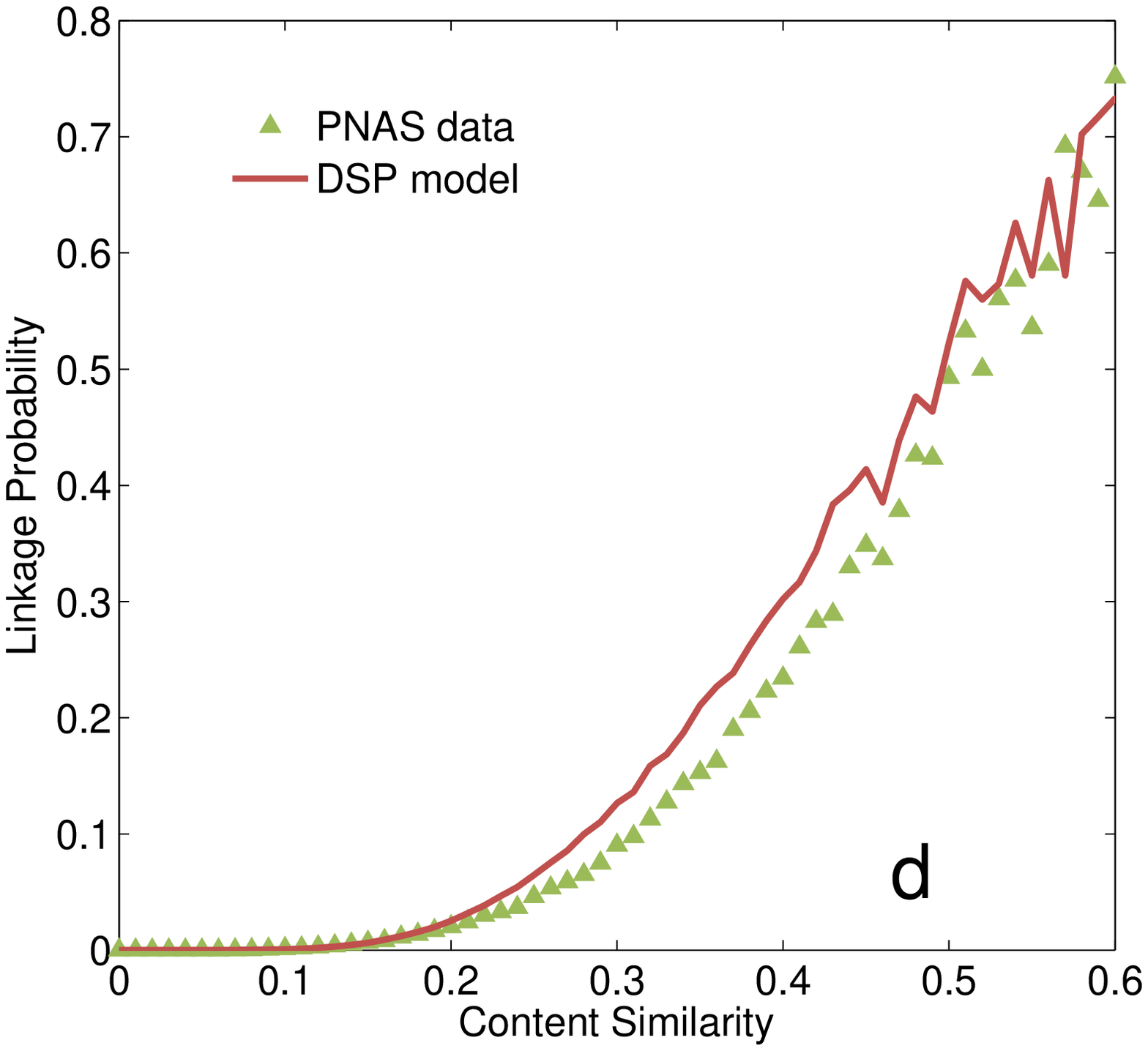}
\caption{\label{fig:DSP-twoNodes} Evaluation of the
degree-similarity product (DSP) model against the WT10g webpage
network and the PNAS citation network: (a) and (b) distribution of
node in-degree; and (c) and (d) linkage probability as a function of
(bilateral) content similarity.}
\end{figure*}

\begin{figure*}
\includegraphics[width=8cm]{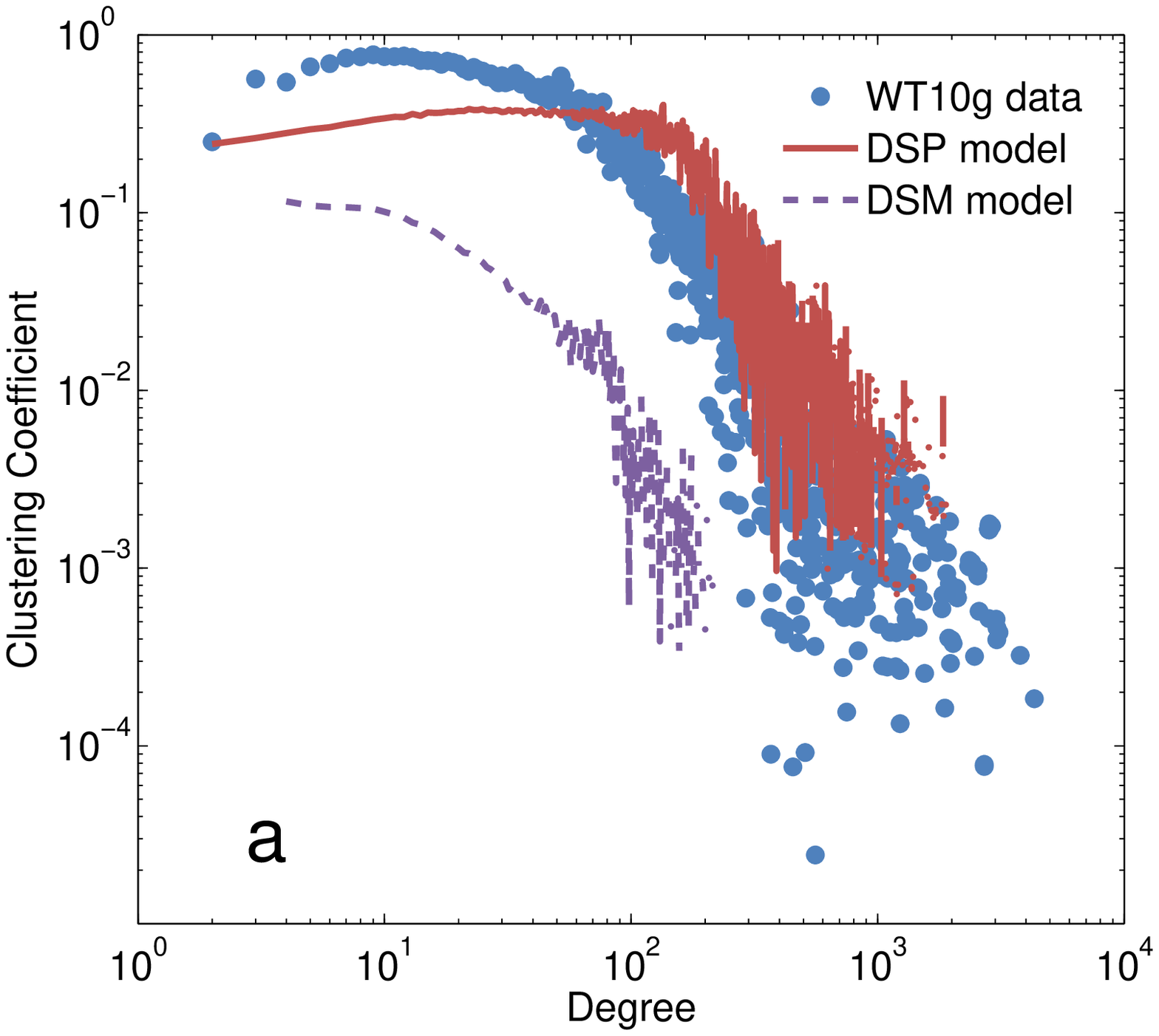}
\includegraphics[width=8cm]{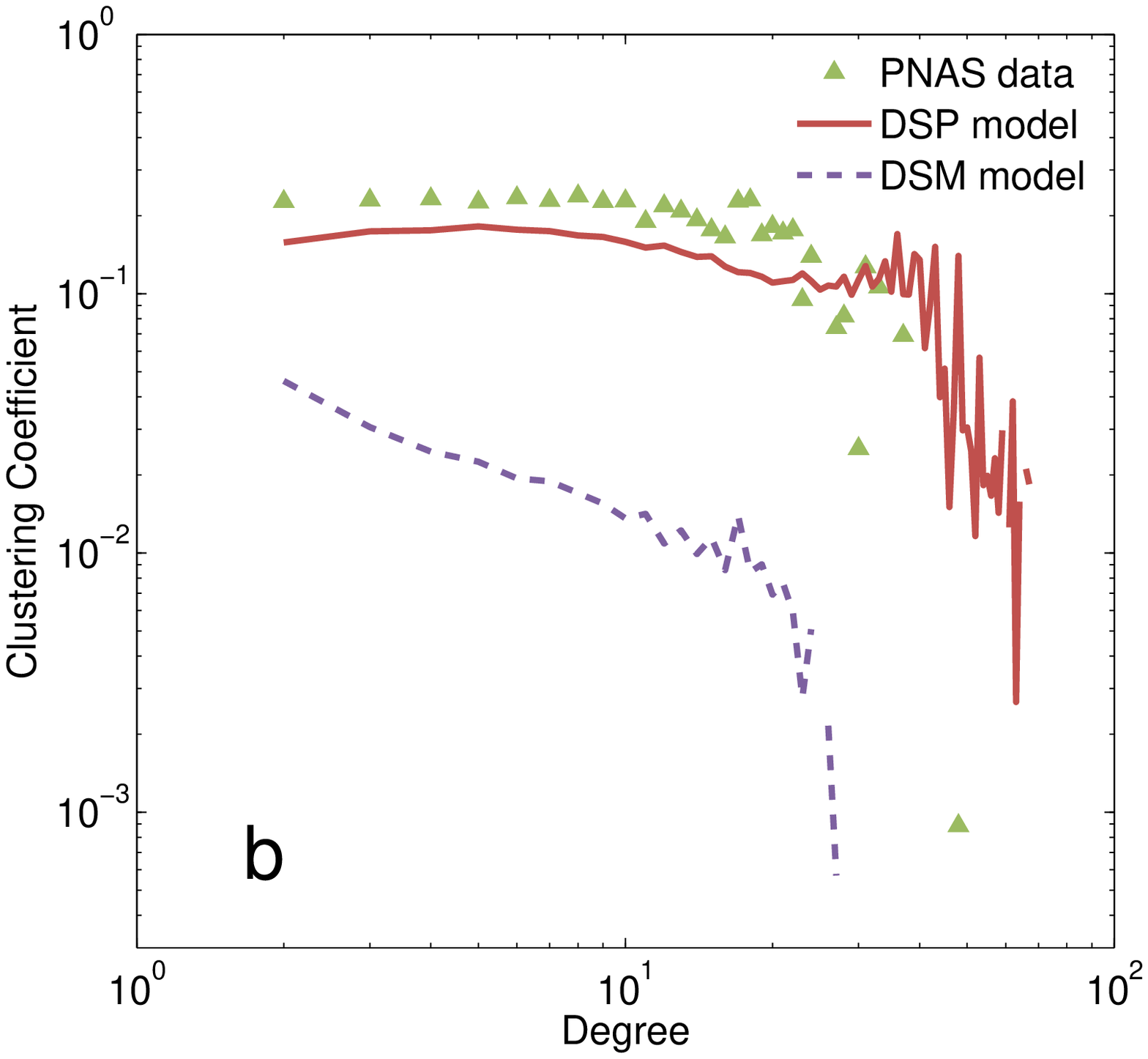}
\includegraphics[width=8cm]{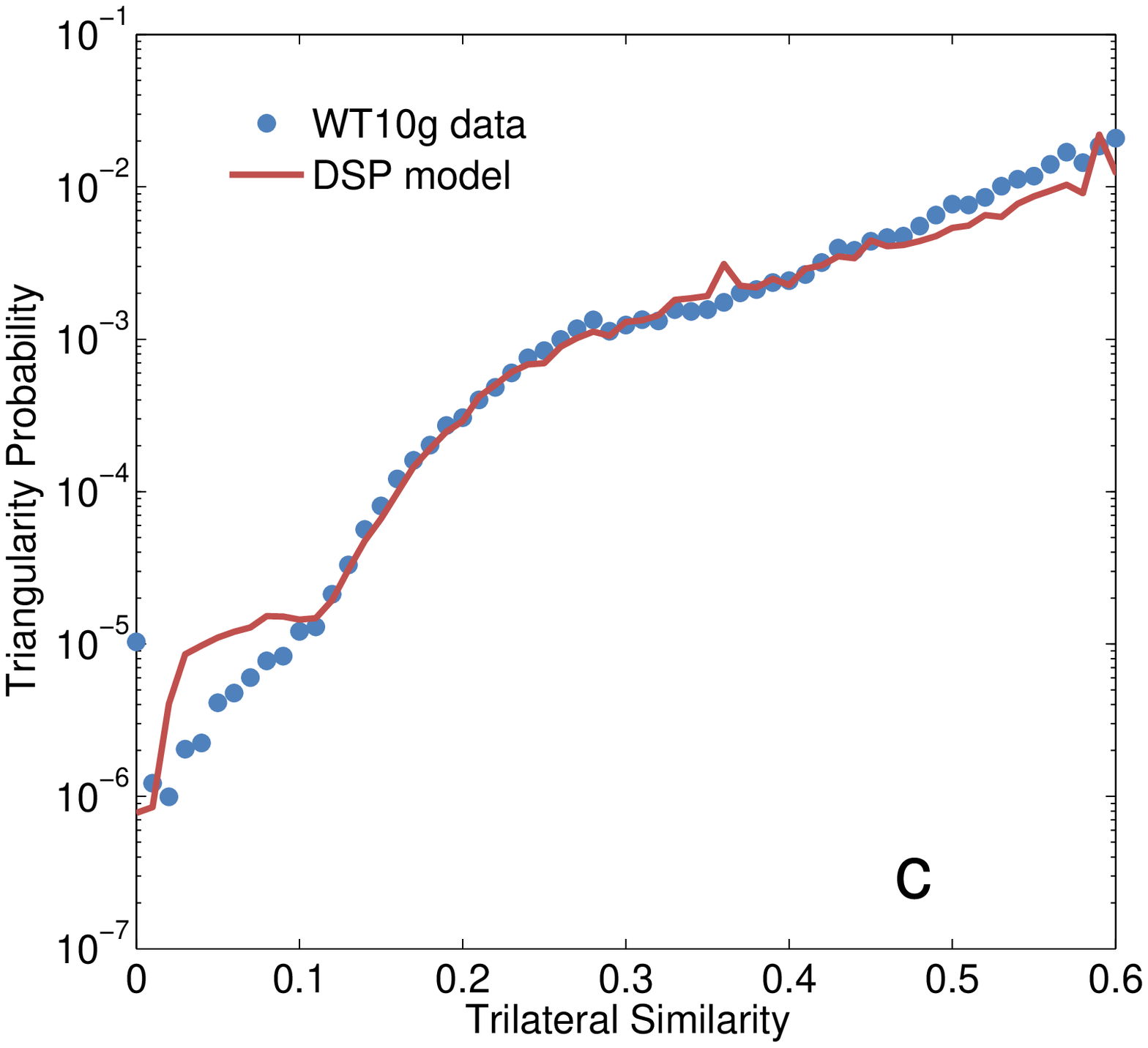}
\includegraphics[width=8cm]{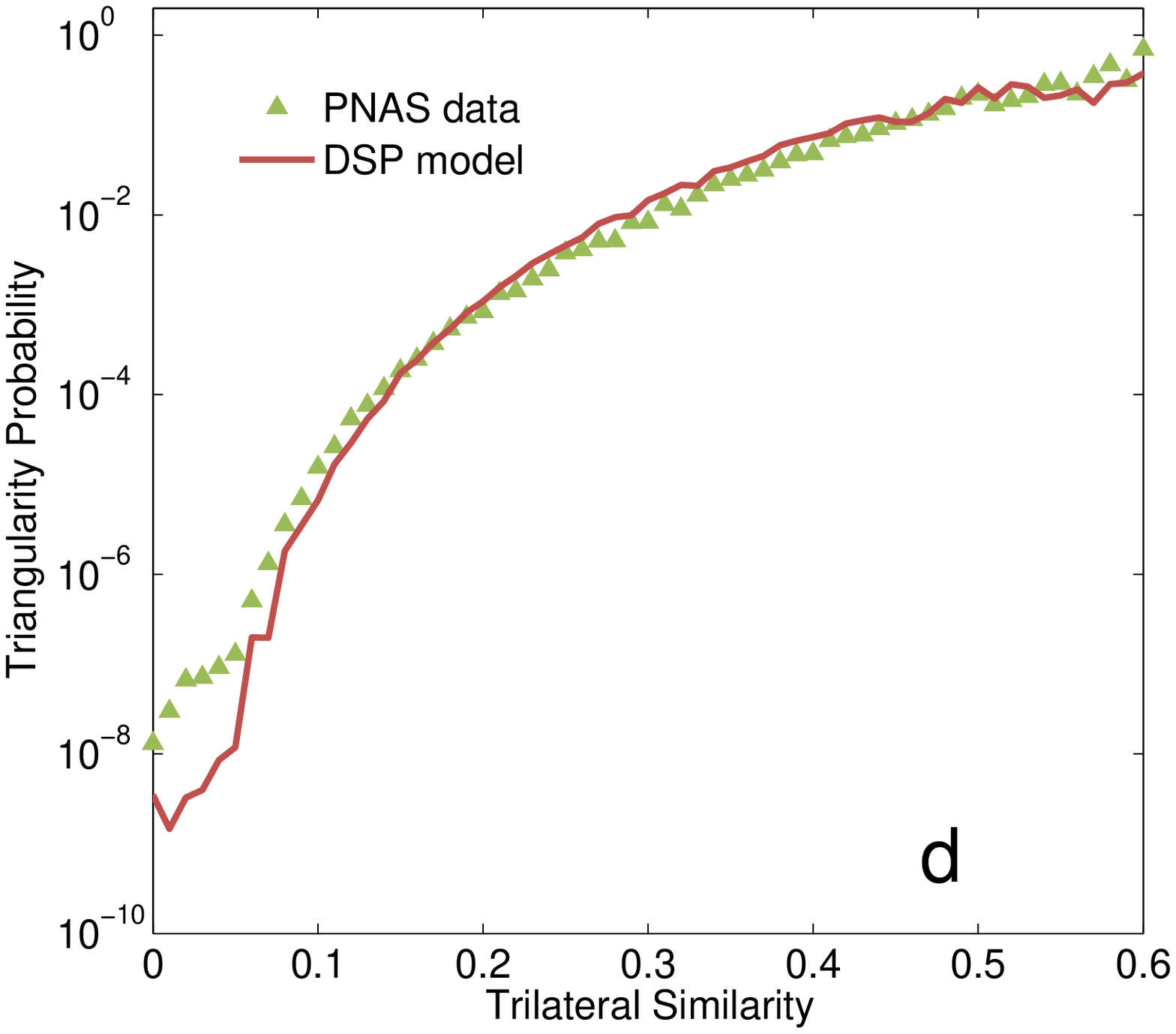}
\caption{\label{fig:DSP-threeNodes} Evaluation of the DSP model
against the two networks: (a) and (b) average clustering coefficient
of $k$-degree nodes; and (c) and (d) triangularity probability as a
function of trilateral (content) similarity.}
\end{figure*}

As shown in Table~\ref{tab:modelEvaluation}, the DSP model well reproduces the number of triangles
and the average clustering coefficient of the two document networks. Figure~\ref{fig:DSP-twoNodes}
and Figure~\ref{fig:DSP-threeNodes} show that the model also closely resembles the two networks'
distribution of node in-degree, linkage probability as a function of content similarity, clustering
coefficient as a function of node degree, and triangularity probability as a function of trilateral
similarity. The average clustering coefficient of nodes with in-degree $k$  (see
Figure~\ref{fig:DSP-threeNodes}(a) and (b)) gives details of a network's triangular clustering
structure.

Table~\ref{tab:parameters} gives the parameters used in the modelling. The value of the parameters
are tuned for best fitting. Our simulation shows that for both the real networks, the best
modelling result is obtained when $\beta_{1}$ (in Equation~\ref{eq:orignal node})  and $\beta_{2}$
(in Equation~\ref{eq:target node}) take different values. This suggests that node out-degree and
in-degree have different weights in choosing the starting and ending nodes of a link.  The values
of $\beta_{1}$ and $\beta_{2}$  for modelling the WT10g data are smaller than those for the PNAS
data. This suggests that a poorly linked webpage has less difficulty in acquiring a new link in
comparison with a poorly cited article. A larger value of $\lambda$ is used for the PNAS data. This
indicates that content similarity plays a relatively stronger role than node connectivity in the
growth of the citation network.

\section{Conclusion}
\label{sec:5}

It is known that document networks show a power-law degree distribution and a positive relation
between the linkage probability and content similarity. In this paper, we show that document
networks also contain  very large numbers of triangles,  high values of clustering coefficient, and
a strong correlation between the triangularity probability and trilateral similarity. These three
properties are not captured by the previous DSM model where a new node tends to link with an old
node which is either popular or similar.

Our intuition is that a link tends to attach between two documents which are both popular and
similar. We propose the degree-similarity product (DSP) model which resembles this behaviour by
using the preferential attachment based on a product function of node connectivity and content
similarity. Our model reproduces all the above topological and content properties with remarkable
accuracy. Our work provides a new insight into the structure and evolution of document networks and
has the potential to facilitate the research on new applications and algorithms on document
networks. Future work will mathematically analyse the DSP model, examine different types of
triangles in document networks, and investigate the possible relation between the triangular
clustering and the formation of communities in document networks.

\section*{Acknowledgements}
This work is supported by the National Key Basic Research Program of
China under grant no.2004CB318109 and the National Natural Science
Foundation of China under grant number 60873245. S.\,Zhou is
supported by the Royal Academy of Engineering and the UK Engineering
and Physical Sciences Research Council (EPSRC) under grant
no.10216/70.


\section*{References}
\begin{thebibliography}{22}


\bibitem {Watts1998} Watts D~J and Strogatz S~H 1998 {\it Nature} {\bf 393} 440
\bibitem {Barab'asi1999} Barab\'asi A-L and Albert R 1999 {\it Science} {\bf 286} 509
\bibitem {Albert2002} Albert R and Barab\'asi A-L 2002 {\it Rev. Mod. Phys.} {\bf
74} 47
\bibitem {Newman2003} Newman M~E~J 2003 {\it SIAM Review} {\bf
45} 167

\bibitem {Kleinberg2001} Kleinberg J. and Lawrence S. 2001
{\it Science} {\bf 294} 1849
\bibitem {Menczer2002} Menczer F 2002 {\it Proc. Natl. Acad. Sci. USA} {\bf
99} 14014

\bibitem {Fenner2006} Fenner T~I, Levene M and Loizou G 2006
{\it ACM Transaction on Internet Technology} {\bf 6} 117

\bibitem {Menczer2004} Menczer F 2004 {\it Proc. Natl. Acad. Sci. USA} {\bf
101} 5261

\bibitem {Cheng2007} Cheng X-Q, Ren F-X, Cao X-B and Ma J. 2007
{\it Proceedings of the IEEE/WIC/ACM International Conference on Web
Intelligence} p~81

\bibitem {Toivonen2006} Toivonen R, Onnela J-P, Saram\"{a}ki J, Hyv\"{o}nen J, and Kaski K 2006
{\it Physica} A {\bf 371} 851

\bibitem {Kumpula2007} Kumpula J~M, Onnela J-P, Saram\"{a}ki J, Kaski K and Kert\'{e}sz J 2007
{\it PRL} {\bf 99} 228701

\bibitem {Soboroff2002} Soboroff I 2002 {\it Proceedings of the 25th annual international ACM SIGIR conference} p~423

\bibitem {Chiang2005} Chiang W-T~M, Hagenbuchner M and Tsoi A~C 2005
{\it Special interest tracks and posters of the 14th international conference on World Wide Web}
p~938

\bibitem
{Schank2005} Schank T and Wagner D 2005 {\it Journal of Graph Algorithms and Applications} {\bf 9}
265
\bibitem {Serrano2006} Serrano M~A and Bogu\~{n}\'{a} M 2005 {\it Phys. Rev.} E {\bf
74} 056114
\bibitem {Fagiolo2007} Fagiolo G 2007 {\it Phys. Rev.} E {\bf
76} 026107
\bibitem {Arenas2008} Arenas A, Fern\'{a}ndez A, Fortunato S and G\'{o}mez S 2008 {\it J. Phys. A: Math. Theor.} {\bf
41} 224001
\bibitem {Zhou2007} Zhou Shi, Cox I~J and Petricek V 2007
{\it Proceedings of the 9th IEEE International Symposium on Web Site Evolution} p~73



\bibitem {Salton1989} Salton G 1989
{\it Automatic text processing: the transformation, analysis, and
retrieval of information by computer} (Addison-Wesley Longman
Publishing Co., Inc.)

\bibitem{Ricardo1999} Ricardo B-Y and Berthier R-N 1999
{\it Modern Information Retrieval} (Addison-Wesley Longman
Publishing Co., Inc.)

\bibitem {Callaway2001} Callaway D~S, Hopcroft J~E, Kleinberg J~M and Newman M~E~J 2001 {\it Phys. Rev.} E {\bf
64} 041902

\bibitem {Krapivsky2002} Krapivsky P~L and Redner S 2002 {\it Computer Networks} {\bf
39} 261



\end{thebibliography}
\end{document}